# MAFcounter: An efficient tool for counting the occurrences of k-mers in MAF files

Michail Patsakis[1,2,+], Kimonas Provatas[1,2,+], Ioannis Mouratidis[1,2,*], Ilias Georgakopoulos-Soares[1,2,*]

[1] Institute for Personalized Medicine, Department of Biochemistry and Molecular Biology, The Pennsylvania State University College of Medicine, Hershey, PA, USA.
[2] Huck Institute of the Life Sciences, Pennsylvania State University, University Park, PA, USA.
[+]These authors contributed equally.
[*] Corresponding authors:  ioannis.mouratidis@psu.edu, izg5139@psu.edu

**Abstract**

**Motivation:** With the rapid expansion of large-scale biological datasets, DNA and protein sequence alignments have become essential for comparative genomics and proteomics. These alignments facilitate the exploration of sequence similarity patterns, providing valuable insights into sequence conservation, evolutionary relationships and for functional analyses. Typically, sequence alignments are stored in formats such as the Multiple Alignment Format (MAF). Counting k-mer occurrences is a crucial task in many computational biology applications, but currently, there is no algorithm designed for k-mer counting in alignment files.

**Results:** We have developed MAFcounter, the first k-mer counter dedicated to alignment files. MAFcounter is multithreaded, fast, and memory efficient, enabling k-mer counting in DNA and protein sequence alignment files.

**Availability:** The MAFcounter package and its Python bindings are released under GPL license as a multi-platform application and are available at:
https://github.com/Georgakopoulos-Soares-lab/MAFcounter

**Contact:** ioannis.mouratidis@psu.edu, izg5139@psu.edu

## Introduction

The rapid progress in high-throughput sequencing technologies has enabled the generation of reference genome and proteome assemblies across organisms from all domains of life and of different individuals across human populations in an ever-increasing pace (Rhie et al. 2021; Darwin Tree of Life Project Consortium 2022; Karczewski et al. 2020; Exposito-Alonso et al. 2020). Genomes and proteomes are composed of nucleotide or peptide sequences, respectively, which can be divided into fixed-length substrings called k-mers, where *k* represents the specified length (Moeckel et al. 2024). The frequency of k-mers varies substantially across genomes and proteomes of different species as well as in genomic sub-compartments (Georgakopoulos-Soares et al. 2021; Chor et al. 2009; Chantzi et al. 2024; Koulouras and Frith 2021).

Counting the occurrence of all k-mers in biological sequences is a crucial step in many bioinformatic applications such as genome assembly, sequence comparison, sequence clustering, error correction of sequencing reads, and genome size estimation (Kelley, Schatz, and Salzberg 2010; Allam, Kalnis, and Solovyev 2015; Salmela and Schröder 2011; Roberts et al. 2004; Williams et al. 2013). K-mers hold significant potential for understanding biological processes, as their occurrences can reveal key aspects of genomic features, including functional elements, repetitive sequences, transcription factor binding sites and variations in the genome, as well as reflect the imprints of DNA damage and repair patterns (di Iulio et al. 2018; Georgakopoulos-Soares et al. 2023; Poulsgaard et al. 2023; Nordström et al. 2013). Proteomes, along with genomes, also play a critical role in k-mer counting analysis by providing insights into coding sequences, functional annotations, structural variations, and evolutionary processes (Bourgeas et al. 2023). Multiple k-mer counting tools have been developed including Jellyfish (Marçais and Kingsford 2011), DSK (Rizk, Lavenier, and Chikhi 2013), Gebril (Erbert, Rechner, and Müller-Hannemann 2017), KMC3 (Kokot, Dlugosz, and Deorowicz 2017) among others, all of which work on a single set of sequences, most often provided in FASTA file format. Nevertheless, there is no method that can count k-mer occurrences in multiple alignment files.

Here, we developed MAFcounter, the first k-mer counter tool for multiple alignment files. MAFcounter takes as input a MAF file and calculates the occurrences of k-mers across each sequence in the alignment, handling both genomic and proteomic sequences effectively within multiple alignment contexts.

## Materials and methods

### Implementation of MAFcounter
MAFCounter is a high-performance multi-threaded command-line tool implemented in C++ with Python bindings distributed through PyPI. It leverages the Google SparseHash library for efficient memory usage in hash-based data structures and uses the moodycamel::ConcurrentQueue for lock-free multi-threaded data processing. For certain configurations which require uint128 bit integers the Boost Multiprecision cpp_int library is employed.. The implementation supports flexible command-line options, including aggregation of reverse-complement k-mers and writing results either as individual genome files or a consolidated output.

## Algorithm Overview

MAFCounter starts by dividing the input MAF file into N chunks to distribute the processing load to N threads. Each chunk is designed to contain approximately the same number of complete alignment blocks, ensuring that each block belongs exclusively to one chunk. After the chunks have been defined the extraction process starts using N producer threads that extract k-mers and N consumer threads that merge the intermediate results while the inter-thread communication is being handled by a lock-free thread safe queue. Producer threads process the sequences of each alignment block within their assigned chunk, using a sliding window of length K to construct a custom data structure called kmerGroups, which are small hash-map structures subsequently pushed to the corresponding queues for consumer threads to merge into the final results. Consumer threads continuously poll the concurrent queue for incoming kmerGroups.The use of kmerGroups, individual hash maps for each alignment block is motivated by the structural properties of alignment blocks, where sequences exhibit significant similarity due to their alignment. This results in repetitive k-mers across sequences within the block. By aggregating k-mers at the block level, updates to the global hashmap are consolidated, minimizing the number of operations required per sequence and instead performing a single update per block. Each k-mer within a kmerGroup is mapped to a specific consumer thread based on a hash function, followed by a modulo operation with the total number of consumers. This ensures that each k-mer is consistently processed by the same consumer thread, eliminating the need for locking and enabling conflict-free concurrent merging, significantly improving performance. In the case of genome MAF files, when the -c flag (canonical or reverse complement k-mer extraction) is enabled, each k-mer is compared with its reverse complement, and the lexicographically smaller of the two is selected as a representative for both k-mers. Upon completion of the k-mer extraction process and the termination of producer and consumer threads, resources are released, and file writing begins. A separate file containing the respective k-mers is created for each genome ID, with the workload being evenly distributed to file writing threads. On the other hand, when the -s or --single_file_output option is specified, the output is consolidated into a single, more compact file grouped first by k-mer and subsequently by genome, to allow for easier comparison of the discrepancies between k-mer frequencies across genomes. In this mode, the writing process operates in a single-threaded manner.

## Counting Strategy

MAFcounter implements an in-memory hash table based strategy to count k-mers, thus memory consumption optimizations are necessary to prevent crashes. To optimize memory usage, MAFcounter leverages the observation that many k-mers, particularly for larger values of k, occur only once per source ID. Initially, k-mers are stored in a Google Dense Hash Set without explicit counts, assuming they are unique. If a k-mer reoccurs its count is stored with an initial value of two. This counting strategy was inspired by KCOSS handling of first-occurrence k-mers (Tang et al. 2022). This approach minimizes the storage overhead for single-occurrence k-mers, effectively reducing memory consumption while accurately maintaining k-mer frequencies, especially in datasets dominated by unique k-mers (singletons) at larger values of k.

**Data Structures and Optimizations**

To optimize memory usage and speed, the tool encodes k-mers using a compact representation where each nucleotide is represented by 2 bits: 'A' as 00, 'C' as 01, 'G' as 10, and 'T' as 11. This encoding allows k-mers to be stored as integers (uint32_t for k ≤ 15 and uint64_t for 16 ≤ k ≤ 31 and uint128_t for 32 ≤ k ≤ 63 ), reducing memory overhead and accelerating computations. For peptide sequences, the tool employs a similar strategy, encoding amino acids in the peptide alphabet using 5 bits per amino acid character. This compact encoding supports efficient storage and computation for peptide k-mers, with data types selected based on k-mer length to balance memory usage and performance. Specifically, uint16_t is used for k=3, uint32_t for 4 ≤ k ≤ 6, uint64_t for 7 ≤k ≤ 12, and uint128_t for 13 ≤ k ≤ 25. This approach ensures reliable and efficient handling of peptide k-mers across a wide range of lengths. To optimize memory usage during extraction, by default genome identifiers are mapped to uint8 integers, limiting the number of unique genomes to 256. This greatly reduces memory consumption while keeping the process efficient. Since most MAF files contain fewer than 100 unique genomes, this tradeoff is practical and works well for the vast majority of cases, with only rare exceptions needing more identifiers. Nevertheless the user has the option to provide the –large_genome_count flag which accounts for larger alignments ramping up the number of unique genomes to 65536 and making the tool future proof. MAFcounter uses Google's Dense Hash Map from the Sparse Hash library to efficiently store k-mer counts. The library offers two variants: Sparse and Dense Hash Maps. Both versions reserve two keys in the key space for optimization, but the Sparse version is more memory-efficient, while the Dense version offers faster lookups. Given MAFcounter's need for frequent lookups, we opted for the Dense Hash Map to prioritize speed over minimal memory usage. Google's Dense Hash Map is not inherently thread-safe, but the need for locking was eliminated by design; the system ensures that the same keys (k-mers) are always assigned to the same consumer thread, preventing concurrent access to the same hashmap entries. These hash maps provide constant-time operations with minimal memory overhead. We optimized the tool by allowing users to customize the k-mer count storage type in the hash map, choosing from options like uint8, uint16, uint32. This is controlled by the max_kmer_count flag, enabling users to save memory when the maximum k-mer count is known in advance or achieve higher accuracy for large counts using the $2^{32}$ option. If an overflow occurs, the program notifies the user in the generated report. The choice of data structures is tailored according to the k-mer length to maximize efficiency, with some optimizations automatically determined by the algorithm and others customizable by the user through specific flags.

**Multithreading**

MAFcounter is designed to run on multiple threads, enabling parallel processing of alignment blocks. This multithreaded approach allows the tool to utilize multi-core processors effectively, reducing processing times for large MAF files. By distributing the workload as well as file writing across multiple threads, MAFcounter can handle extensive datasets more efficiently, enhancing its scalability and performance. MAFcounter utilizes a lock-free queue for inter-thread communication and a mapping of the same keys to the same consumers to prevent hash-map key collisions. Locking is applied only to minimal sections of the code, such as during the mapping of genome names to uint8 integers for memory efficiency, which occurs at most 256 times. Performance benchmarks were carried out on selected regions of the human genome alignment totalling 2 GB, to assess the efficiency of MAFcounter across varying k-mer lengths. The benchmarks also analyzed how performance scales with the number of CPU cores, identifying the point where adding more cores no longer improves performance significantly. As shown in Figure 1, smaller k-mers take longer to reach their optimal thread count. This aligns with the theoretical model, as smaller k-mers result in hash maps with fewer keys but larger counts per key thus resulting in less time spent writing new keys and files.

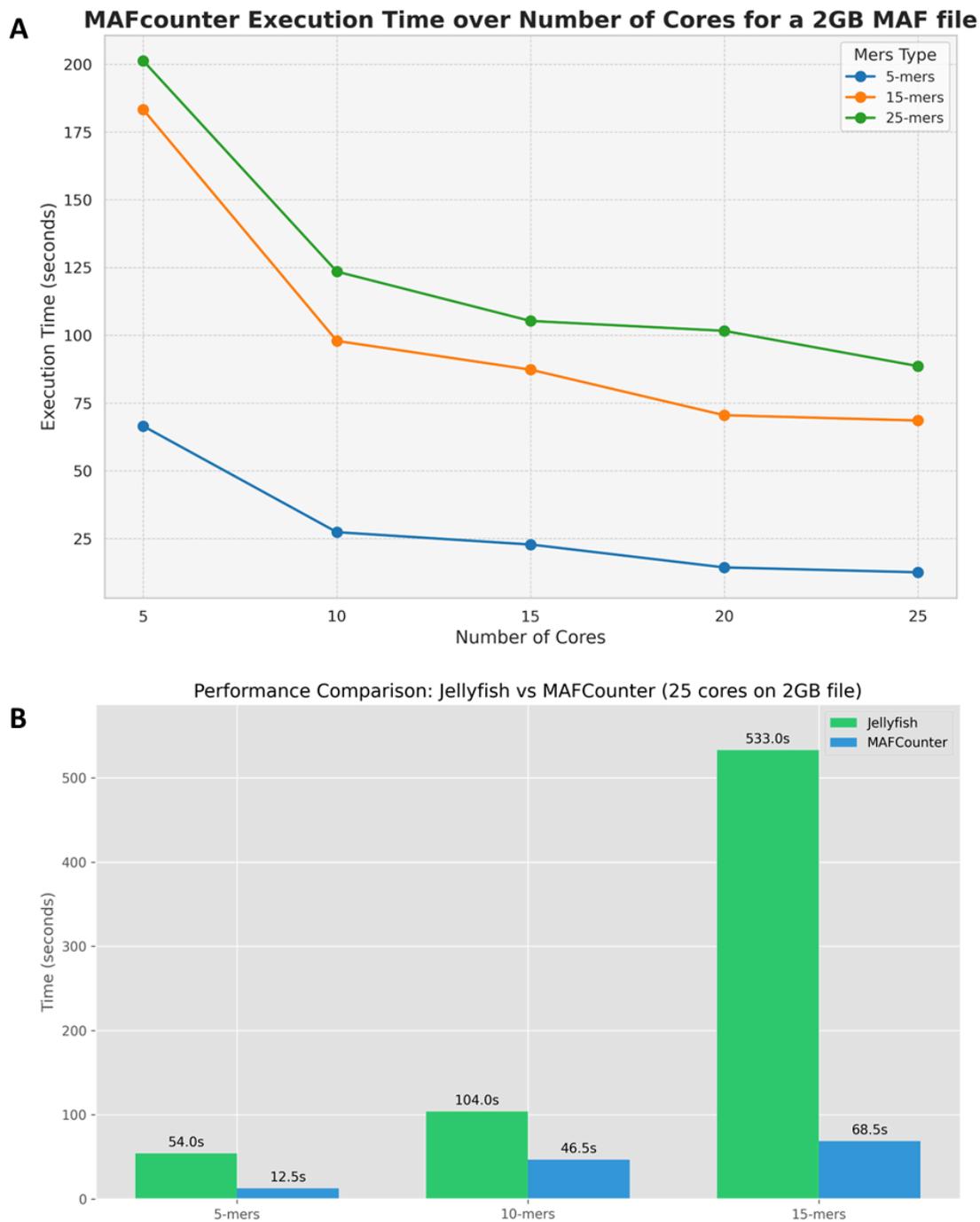

**Figure 1: Benchmarks on MAFcounter and comparison with Jellyfish . A.** Performance Benchmark of MAFcounter on 2GByte file for k=5,15,25 and number of cores from 5 to 25 with step 5. **B.** Performance comparison of MAFcounter with Jellyfish on 25 threads. The MAF file was split to multiple fasta files before running Jellyfish.

**Examples**

Extract 15mers on both strands from input.maf using 16 cores and write the results in a single file kmers_txt:

**./maf_counter -c -s 15 input.maf 16**

Gives an output file kmer_counts.txt inside results_counter directory with the following format

<SEQUENCE_1> <GENOME_ID1>:<COUNT_1>,<GENOME_ID2>:<COUNT_2>

<SEQUENCE_2> <GENOME_ID1>:<COUNT_3>,<GENOME_ID2>:<COUNT_4>

And the -s flag is omitted

**./maf_counter -c 15 input.maf 16**

Results in multiple_files , one per genome_id inside the results_counter directory with the following format:

<SEQUENCE_1> <COUNT_1>

<SEQUENCE_2> <COUNT_2>

….

<SEQUENCE_N> <COUNT_N>

**Discussion**
In this study, we developed MAFcounter, a robust counter tool designed to efficiently analyze sequence alignments and quantify the occurrence of k-mers. MAFcounter provides a valuable resource for researchers working on comparative genomics, evolutionary biology, and other fields where sequence alignment is critical and counting k-mer occurrences in a first step in data interpretation. By enabling rapid and accurate counting of k-mers in multiple alignments, MAFcounter simplifies large-scale genomic analyses and reduces the computational burden associated with handling extensive datasets. The program's modularity and scalability allow for its integration into existing pipelines, enhancing its applicability to a wide range of research projects, from basic biology to clinical genomics.

**Code Availability**
The GitHub code and all the related material is provided at:
https://github.com/Georgakopoulos-Soares-lab/MAFcounter

**Competing interests**
No competing interest is declared.

**Funding**
Research reported in this publication was supported by the National Institute of General Medical Sciences of the National Institutes of Health under Award Number R35GM155468. The content is solely the responsibility of the authors and does not necessarily represent the official views of the National Institutes of Health.